\documentclass[traditabstract]{aa}
\usepackage{graphicx}

\usepackage{bm}
\usepackage{epsfig}
\usepackage[colorlinks=true,linkcolor=blue]{hyperref}

\newcommand{\be}{\begin{equation}}
\newcommand{\ee}{\end{equation}}

\newcommand{\bea}{\begin{eqnarray}}
\newcommand{\eea}{\end{eqnarray}}

\newcommand{\Gm}{\Gamma}
\newcommand{\dl}{\delta}
\newcommand{\Dl}{\Delta}

\newcommand{\et}{\eta}

\newcommand{\kp}{\kappa}
\newcommand{\lm}{\lambda}
\newcommand{\Lm}{\Lambda}

\newcommand{\rh}{\rho}

\newcommand{\Om}{\Omega}

\newcommand{\Rarrow}{\Rightarrow}

\newcommand{\nn}{\nonumber}

\newcommand{\varep}{\varepsilon}

\begin{document}

\title{A conventional approach to the dark-energy concept}

\author{K. Kleidis \inst{1}
  \and
  N. K. Spyrou \inst{2}}

\institute{Department of Mechanical Engineering, Technological Education Institute of Serres, 621.24 Serres, Greece \\e-mail: \texttt{kleidis@teiser.gr} \and Department of Astronomy, Aristoteleion University of Thessaloniki, 541.24 Thessaloniki, Greece \\e-mail: \texttt{spyrou@astro.auth.gr}}

\offprints{N. K. Spyrou}

\date{Received ..... ; accepted .....}

\titlerunning{A conventional approach to the dark-energy concept}

\authorrunning{K. Kleidis \& N. K. Spyrou}

\abstract{Motivated by results implying that the constituents of dark matter (DM) might be collisional, we consider a cosmological (toy-) model, in which the DM itself possesses some sort of thermodynamic properties. In this case, not only can the matter content of the Universe (the baryonic component, which is tightly gravitationally-bounded to the dark one, also being included) be treated as a classical gravitating fluid of positive pressure, but, together with all its other physical characteristics, the energy of this fluid's internal motions should be taken into account as a source of the universal gravitational field. In principle, this form of energy can compensate for the extra (dark) energy, needed to compromise spatial flatness, while the post-recombination Universe remains ever-decelerating. What is more interesting, is that, at the same time (i.e., in the context of the collisional-DM approach), the theoretical curve representing the distance modulus as a function of the cosmological redshift, $\mu (z)$, fits the Hubble diagram of a multi-used sample of supernova Ia events quite accurately. A cosmological model filled with collisional DM could accommodate the majority of the currently-available observational data (including, also, those from baryon acoustic oscillations), without the need for either any dark energy (DE) or the cosmological constant. However, as we demonstrate, this is not the case for someone who, although living in a Universe filled with self-interacting DM, insists on adopting the traditional, collisionless-DM approach. From the point of view of this observer, the cosmologically-distant light-emitting sources seem to lie farther (i.e., they appear to be dimmer) than expected, while the Universe appears to be either accelerating or decelerating, depending on the value of the cosmological redshift. This picture, which, nowadays, represents the common perception in observational cosmology, acquires a more conventional interpretation within the context of the collisional-DM approach.}

\keywords{Accelerated Expansion -- Dark Matter -- Dark Energy}


\maketitle

\section{Introduction}

The beginning of the $21^{st}$ century was one of the most exciting epochs for cosmology as a science. According to observational data on the temperature variations in the cosmic microwave background (CMB) that became in public at that epoch (de Bernardis et al. 2000; Jaffe et al. 2001; Padin et al. 2001; Stompor et al. 2001; Netterfield et al. 2002), now, we are quite confident that the Universe can be adequately described by a spatially flat Robertson-Walker (RW) cosmological model. 

As a consequence, the total energy density, $\varep$, of the Universe matter-energy content, in units of the energy density $\varep_c = \rh_c c^2$ (equivalent to the critical rest-mass density, $\rh_c= \frac{3 H_0^2}{8 \pi G}$, where $H_0$ is the Hubble parameter at the present epoch, $c$ is the velocity of light, and $G$ is Newton's universal constant of gravitation), should be very close to unity, $\Om = \frac{\varep}{ \varep_c} \simeq 1$, i.e., much larger than the measured quantity, $\Om_M = \frac{\rh}{\rh_c} \simeq 0.3$ (Komatsu at al. 2009). 

At the same time, high-precision distance measurements, performed with the aid of the supernovae Ia (SNe Ia) standard candles, indicated that, in any cosmological model with vanishing cosmological constant, $\Lm$, the far-off light-emitting sources appear to be dimmer than expected (Riess et al. 1998; Perlmutter et al. 1999). 

The observational data then seemed to favour a Universe of collisionless content (i.e., filled with matter in the form of dust) and $\Lm \neq 0$, in which $\Om_M \simeq 0.3$ and $\Om_{\Lm} = \frac{\Lm c^2}{3 H_0^2} \simeq 0.7$ (Riess et al. 2001, 2004). Since a non-vanishing cosmological constant (necessarily) involves a repulsive (gravitational) force (see, e.g., Sahni 2004), the apparent dimming of the distant light-emitting sources was attributed to a relatively recent phase of accelerated expansion (see, e.g., Linder 2008). 

The onset of the dimming of the cosmologically distant indicators used (hence, the associated transition from acceleration to deceleration), takes place at a relatively low value of the cosmological redshift, $z$, the so-called transition redshift, $z_t$, which, nowadays, is being (observationally) set at $z_t = 0.46 \pm 0.13$ (Riess et al. 2004, 2007). In this case, the cosmological constant can be determined observationally, since, on theoretical grounds, $z_t = \left (2 \frac{\Om_{\Lm}} {\Om_M} \right )^{1/3} - 1$ (see, e.g., Perivolaropoulos 2007, Eq. (21)), i.e., the transition redshift depends on the value of $\Lm$. The particle-physics vacuum does contribute an effective cosmological constant, which could serve (also) as compensation to the extra energy needed to flatten the Universe (Sahni \& Starobinsky 2000). Unfortunately, the energy-density attributed to such a source is $10^{123}$ times larger than what is observed (see, e.g., Padmanabhan 2003; Sahni 2004). 

Hence, it became evident that, for the above-mentioned observational results to be reconciled within a unified theoretical framework, a different approach (i.e., other than the cosmological constant) was needed.

In this context, theorists have focused on two main areas: \texttt{(i)} The introduction of an exotic, negative-pressure fluid, the dark energy (Caldwell et al. 1998), occasionally referred to as quintessence (Carroll 1998), and \texttt{(ii)} the consideration of alternative-gravity theories, such as the scalar-tensor theories (Esposito-Farese \& Polarski 2001) and the $f(R)$-gravity (Capozziello et al. 2003), together with several braneworld scenarios, including DGP-gravity (Dvali et al. 2000) and the landscape scenario (Bousso \& Polchinski 2000). 

Other physically-motivated models, predicting an accelerated expansion, have also appeared in the literature, involving holographic gravity (Cohen et al. 1999; Li 2004; Pav\'{o}n \& Zimdahl 2005), Chaplygin gas (Kamenshchik et al. 2001; Bean \& Dor\'{e} 2003; Sen \& Scherrer 2005), Cardassian cosmology (Freese \& Lewis 2002; Wang et al. 2003), theories of compactified internal dimensions (Perivolaropoulos 2003), and mass-varying neutrinos (Fardon et al. 2004; Peccei 2005). 

However, most of these attempts were inhibited by the so-called coincidence problem, i.e., the need to explain why the Universe transferred from deceleration to acceleration so recently (see, e.g., Perivolaropoulos 2007).

It has been more than a decade since the first observations, which provoked the aforementioned scientific (r)evolution, became in public, and, still, no undisputed theoretical framework has been developed to accommodate them (for a detailed review see, e.g., Caldwell \& Kamionkowski 2009). In the meantime, the need for an extra (dark) energy component has been confirmed by other observational methods, including galaxy clusters dynamics (Allen et al. 2004), the integrated Sachs-Wolfe (ISW) effect (Boughn \& Crittenden 2004), and baryon acoustic oscillations (BAO) (Eisenstein et al. 2005; Percival et al. 2010). In spite of the wealth of references, previous studies have been far from exhaustive (see, e.g., Albrecht et al. 2006; Peacock et al. 2006) and, perhaps, we should keep our options open, also, to more conventional interpretations (see, e.g., Buchert 2000, 2001; Kolb et al. 2006; Celerier 2007; Ellis 2009). 

On the other hand, much evidence has already been accumulated in support of a (non-baryonic) DM component in the Universe matter-content (see, e.g., Tegmark et al. 2006; Spergel et al. 2007). Among other pieces of evidence, this support includes flattened galactic rotation curves (Begeman et al. 1991; Borriello \& Salucci 2001), the weak gravitational lensing of distant galaxies by (some dark) foreground structure (Hoekstra et al. 2002), and the weak modulation of strong lensing around individual massive elliptical galaxies (Moustakas \& Metcalf 2003). On the scale of galaxy clusters, observations (of radial velocities, weak lensing, and X-ray emission) indicate a total mass density almost ten times higher than the corresponding density in baryons (Bahcall \& Fan 1998; Kashlinsky 1998; Tyson et al. 1998). On cosmological scales, the anisotropies observed in the CMB have led to an estimate of the total-mass density of the order of $\Om_M h^2 = 0.1358 \pm 0.0037$, where $h$ is the Hubble parameter in units of $100 \; km/sec$ per $Mpc$ (Komatsu et al. 2009). In contrast, measurements of the light-chemicals' abundances (Olive et al. 2000) have led to an estimate of the baryonic-mass density of the order of $\Om_B h^2 = 0.02273 \pm 0.00062$. The combination of these results suggests that, more than $85 \%$ (by mass) of the matter in the Universe consists of non-luminous and non-baryonic material. 

Although we do not know for certain how the DM came to be formed, a sizeable relic abundance of weakly-interacting massive particles (WIMPs) is generally expected to have been produced as a by-product of the Universe's hot youth (see, e.g., Kolb \& Turner 1990, p. 369). These particles decouple from radiation much earlier than pure-baryonic matter does. Hence, very soon after recombination $(t_R)$, the baryons fall into deep potential wells of the already evolved DM-perturbations and become bounded to them, i.e., for $t > t_R$, there are no freely-floating baryons around (Olive 2003).

Among the various candidates for DM constituents, the thermal WIMPs remain one of the most attractive. They appear, generically, in theories of weak-scale physics beyond the standard model, while giving the appropriate relic abundance (Srednicki et al. 1988; Gondolo \& Gelmini 1991). These particles are also helpful to consider in terms of direct and indirect detection of DM (see, e.g., Jungman et al. 1996; Bertone et al. 2005), because they must have some connection to standard-model particles (see, e.g., Hooper 2009). In addition to debating their precise nature, the scientific community used to argue that the WIMPs should be collisionless. 

However, many results from high-energy particle detectors, such as the ATIC (Chang et al. 2008) and PAMELA (Adriani et al. 2009), combined with data from the Wilkinson microwave anisotropy probe (WMAP) survey (Hooper et al. 2007), have revealed an unusually high electron - positron production in the Universe, much more than anticipated by SNe explosions or cosmic-ray collisions. These results have led many scientists to argue that among the best candidate sources of these high-energy events are the annihilations of WIMPs (see, e.g., Barger et al. 2008; Bergstrom et al. 2008; Cirelli \& Strumia 2008; Regis \& Ullio 2008; Baushev 2009; Cholis et al. 2009a, 2009b; Fornasa et al. 2009; Fox \& Poppitz 2009; Kane et al. 2009; Zurek 2009, for an extensive, though incomplete list), i.e., that the DM constituents can be slightly collisional (see, e.g., Spergel \& Steinhardt 2000; Arkani-Hamed et al. 2009; Cirelli et al. 2009; Cohen \& Zurek 2010), although, some studies disagree with this interpretation (see, e.g., Feng et al. 2010). 

A cosmological model filled with self-interacting DM could be a relatively inexpensive solution to the DE problem, and, several ways of accommodating both the DM and the DE into a unified theoretical framework have been considered (see, e.g., Zimdahl et al. 2001; Bili\'{c} et al. 2002; Balakin et al. 2003; Scherrer 2004; Lima et al. 2008; Basilakos \& Plionis 2009, 2010; Dutta \& Scherrer 2010). In this context, we suggest that, phenomenologically, the self-interacting DM could attribute to the Universe matter-content some sort of fluid-like properties, and (so) lead to a conventional approach to the DE concept.

The main outstanding problem of the current cosmological picture is that the Universe must contain an amount of energy that is considerably higher than the equivalent of the total rest-mass of its matter content. However, if the DM constituents collided with each other frequently enough, enabling their (kinetic) energy to be re-distributed, i.e., if the DM itself possessed some sort of thermodynamic properties, a conventional extra-energy component might be present in the Universe, given by the energy of the internal motions of the collisional-DM fluid.

On this basis, it is worth examining the evolution and the dynamical characteristics of a cosmological model (not necessarily reflecting our own Universe), in which (in principle) there is no DE at all. The matter-energy content of this model consists solely of two components, i.e., the DM (dominant) and the baryonic one (subdominant), both having the abundances attributed to them by analyses of the five-year survey of the WMAP (Dunkley et al. 2008; Komatsu et al. 2009). Accordingly, we demonstrate that these two components are (by themselves) sufficient \texttt{(i)} to reproduce the result that (today) $\Om = 1$, \texttt{(ii)} to account for the observed dimming of the distant light-emitting sources, and \texttt{(iii)} to explain the apparent accelerated expansion of the Universe. All the above provided that, macroscopically, these two constituents (basically the dark one) form a gravitating fluid with a thermodynamical content. In this case, together with all the other physical characteristics, the energy of this fluid's internal motions should (also) be taken into account as a source of the universal gravitational field. Although speculative, the idea that the extra (dark) energy needed to flatten the Universe could be attributed to the internal motions of a collisional-DM fluid is (at least) intriguing.

This paper is organized as follows. In Sect. 2, we explore the dynamical characteristics of a cosmological model driven by an ideal fluid, consisting (mainly) of (thermodynamically-involved) DM, with positive pressure, the volume elements of which perform adiabatic flows. Accordingly, after deriving the corresponding scale factor, we determine the functional form of several parameters of cosmological significance, each one depending on the cosmological redshift, such as the luminosity distance and the distance modulus of the cosmologically-distant light-emitting sources, together with the Hubble and the deceleration parameters, which characterize the cosmic expansion. The corresponding results suggest that, in the context of the collisional-DM treatment, the extra (dark) energy (needed to compromise spatial flatness) can be compensated by the energy of the internal motions of this fluid, while, the post-recombination Universe remains ever-decelerating. However, as we demonstrate in Sect. 3, this is not the case for someone who (although living in a Universe filled with collisional DM) insists on adopting the traditional (collisionless-DM) approach. From the point of view of this observer, besides the need for an extra-energy component (for confronting the CMB-based observational results), the cosmologically-distant light-emitting sources seem to lie farther (i.e., they appear to be dimmer) than expected, while the Universe appears to be either accelerating or decelerating, depending on the value of the cosmological redshift. Finally, we conclude in Sect. 4.

\section{A Universe filled with collisional dark matter}

It is generally accepted that the study of the CMB has proven to be a powerful tool in exploring the post-recombination Universe. According to the various CMB-oriented observational data, the Universe has emerged out of the radiation epoch as a spatially-flat RW model (see, e.g., de Bernardis et al. 2000) \be ds^2 = S^2 (\et) \left [ c^2 d \et^2 - \left ( dx^2 + dy^2 + dz^2 \right ) \right ] \: , \ee where $\et$ is the conformal time and $S (\et)$ is the scale factor. As a consequence, the value of the Hubble parameter at the present epoch is, by definition, given by \be H_0^2 = \frac{8 \pi G}{3} \rh_c \ee (see, e.g., Peacock 1999, p. 77). The evolution of this model depends on the nature of the source that drives the universal gravitational field, i.e., its matter-energy content. 

Along the lines of the collisional-DM approach, in specifying the Universe matter-energy content, we assume that, in principle, there is no DE at all. Instead, we admit that the DM, together with the small, baryonic "contamination" (the latter is too tightly gravitationally-bounded to the former), possess fluid-like properties. In this sense, the collisions of the WIMPs maintain a tight coupling between them and their energy can be re-distributed, i.e., the DM itself also possesses some sort of thermodynamical content. In this case, the evolution of the post-recombination Universe is no longer driven by dust, but, by a gravitating fluid of positive pressure, $p$, satisfying the equation of state \be p = w \rh c^2  \: , \ee where $\rh$ is the rest-mass density (the part, equivalent to the energy density $\rh c^2$, that remains unaffected by the internal motions of the cosmic fluid) and $0 \leq w = \left ( \frac{c_s}{c} \right )^2 \leq 1$ is a dimensionless constant, which measures the square of the speed of sound, $c_s$, in units of $c^2$. Now, the fundamental units of the Universe matter-content are the volume elements of the collisional-DM fluid (elements of fluid, each one consisting always of the same particles).

The motions of the volume elements in the interior of a continuous medium are governed by the equations \be T_{\; \; ; \nu}^{\mu \nu} = 0 \; , \ee where Greek indices refer to the four-dimensional space-time (in connection, Latin indices refer to the three-dimensional spatial slices), the semicolon denotes covariant derivative, and $T^{\mu \nu}$ is the energy-momentum tensor of the Universe matter-content, i.e., basically (but not solely), of the collisional-DM fluid. 

Confining ourselves to the particular case of a perfect fluid, $T^{\mu \nu}$ takes on the standard form \be T^{\mu \nu} = (\varep + p)u^{\mu} u^{\nu} - p g^{\mu \nu} \: , \ee where $u^{\mu} = dx^{\mu}/ds$ is the four-velocity $\left ( u_{\mu}u^{\mu} = 1 \right )$ at the position of a fluid's volume element, $g^{\mu \nu}$ are the contravariant components of the Universe metric tensor, and $\varep$ is this fluid's total-energy density. In an (ideal) equilibrium state, i.e., in the absence of shear, viscocity and heat conductivity, $\varep$ is decomposed to \be \varep = \rh c^2 + \rh \Pi \ee (for a detailed analysis see, e.g., Fock 1959, pp. 81 - 83 and 91 - 94), where $\Pi$ is the potential energy per unit rest-mass, associated with the infinitesimal deformations (expansions or/and compressions) of the fluid. Upon consideration of adiabatic processes, $\Pi$ coincides with the energy of this fluid's internal motions (per unit rest-mass), thus defining $\rh \Pi$ as the corresponding energy density, associated with the thermodynamical content of the (collisional) DM.

Along these lines, the equations represented by Eq. (4) are the hydrodynamic flows of the volume elements in the interior of a perfect-fluid source \be u_{\mu ; \nu} u^{\nu} = \frac{1}{\varep + p} p_{, \kp} \left (\dl_{\mu}^{\kp} - u_{\mu} u^{\kp} \right ) \: , \ee where the comma denotes a partial derivative and $\dl_{\mu}^{\kp}$ is the Kronecker symbol. The equations given in Eq. (7) can be cast in the more convenient form \be \frac{d u^{\kp}}{ds} + \Gm_{\mu \nu}^{\kp} u^{\mu} u^{\nu} = \frac{1}{\varep + p} h^{\kp \lm} p_{, \lm} \: , \ee where $\Gm_{\mu \nu}^{\kp}$ are the Christoffel symbols corresponding to the Universe metric tensor, $g_{\mu \nu}$, and $h^{\kp \lm} = g^{\kp \lm} - u^{\kp} u^{\lm}$ is the projection operator. 

However, in a maximally symmetric cosmological setup, there is no real difference between hydrodynamic flows and the ballistic motions along the (non-intersecting) geodesic trajectories $x^{\mu} = constant$, since, the equations of motion given by Eq. (8) are trivially satisfied by the fluid's volume elements with $u^{\mu} = (1, 0, 0, 0)$. In other words, in comoving coordinates, both sides of Eq. (8) vanish identically, resulting in the geodesic equations \be \frac{d u^{\kp}}{ds} + \Gm_{\mu \nu}^{\kp} u^{\mu} u^{\nu} = 0 \ee and the equations of the ballistic trajectories \be h^{\kp \lm} p_{, \lm} = 0 \: . \ee Notice however that, in this case, Eq. (10) is satisfied only because of the form of the metric tensor in Eq. (1), i.e., even if $p_{, \: \lm}$ does not vanish. Hence, in comoving coordinates, the geodesic motions and the hydrodynamic flows of the cosmological model given by Eq. (1) are, practically, indistinguishable. Therefore, a comoving observer of the cosmic expansion also traces the hydrodynamic flow of the homogeneous cosmic fluid and the Weyl's postulate is valid (see, e.g., Narlikar 1983, p. 91). 

As a consequence, the dynamical evolution of the model given by Eq. (1) is governed by the Friedmann equation (with $\Lm = 0$) of the classical Friedmann-Robertson-Walker (FRW) cosmology \be H^2 = \frac{8 \pi G}{3 c^2} \varep \: , \ee where \be H = \frac{S^{\prime}}{S^2} \ee is the Hubble parameter as a function of the scale factor, and the prime denotes differentiation with respect to $\et$. 

Nevertheless, inherently, there is an essential difference between our model and the rest of the classical FRW cosmologies. In our case, the basic matter constituents (although they may resemble test particles receding from each other) are the volume elements of a collisional-DM fluid, i.e., they possess some sort of internal structure, hence thermodynamical content. Therefore, the functional form of $\varep$ in Eq. (11) is no longer given by $\rh c^2$ alone, but by Eq. (6) (see also Narlikar 1983, pp. 61, 62).

In this model, the first law of thermodynamics for adiabatic flows, given by, \be d \Pi + p d \left ( \frac{1}{\rh} \right ) = 0 \ee (see, e.g., Chandrasekhar 1965), results in \be \Pi = \Pi_0 + wc^2 \ln \left ( \frac{\rh}{\rh_0} \right ) \: , \ee where the constants $\rh_0$ and $\Pi_0$ are assumed to denote the corresponding present-time values. Accordingly, the total-energy density of the Universe matter-energy content is written in the form \be \varep = \rh c^2 \left [ 1 + \frac{\Pi_0}{c^2} + w \ln \left ( \frac{\rh}{\rh_0} \right ) \right ] \: . \ee 

On the other hand, for every value of $w$, the conservation law $T_{\; ; \nu}^{0 \nu} = 0$, in terms of the metric tensor of Eq. (1), yields \be \varep^{\prime} + 3 \frac{S^{\prime}}{S} (\varep + p) = 0 \: , \ee which, upon consideration of Eqs. (3) and (6), results in \be \rh = \rh_0 \left ( \frac{S_0}{S} \right )^3 \: , \ee where $S_0$ is the value of $S (\et)$ at the present epoch. Equation (17) represents the conservation of the total mass in a cosmological model in which matter dominates, i.e., for every $\et$ within the post-recombination epoch (see, e.g., Tsagas et al. 2008).

With the aid of Eqs. (15) and (17), Eq. (11) is written in the form \be H^2 = \frac{8 \pi G}{3} \rh_0 \left ( \frac{S_0}{S} \right )^3 \left [ 1 + \frac{\Pi_0}{c^2} + 3 w \ln \left ( \frac{S_0}{S} \right ) \right ] \: . \ee Now, combining Eqs. (2) and (18), we obtain \be \left ( \frac{H}{H_0} \right )^2 = \Om_M \left ( \frac{S_0}{S} \right )^3 \left [ 1 + \frac{\Pi_0}{c^2} + 3 w \ln \left ( \frac{S_0}{S} \right ) \right ] \: . \ee At the present epoch, where $S = S_0$ and $H = H_0$, we have \be \Om_M \left (1 + \frac{\Pi_0}{c^2} \right ) = 1 \: , \ee from which, the present-time value of the internal energy per unit rest-mass, $\Pi_0$, emerges as \be \Pi_0 = \left ( \frac{1}{\Om_M} - 1 \right ) c^2 \: . \ee Since $\Om_M < 1$, Eq. (21) suggests that, at the present epoch, the energy density, $\rh_0 \Pi_0$, of the internal motions of a gravitating perfect fluid (consisting, mainly, of collisional DM and a small baryonic contamination) dominates over the corresponding rest-mass quantity, i.e., \be \rh_0 \Pi_0 = \left ( \frac{1}{\Om_M} - 1 \right ) \rh_0 c^2 > \rh_0 c^2 \: . \ee But, what is more important, is that, at the same time, the combination of Eqs. (6) and (21) results in the following value of the total-energy density parameter \be \Om =  \frac{\varep_0}{\varep_c} = \frac{\rh_0 c^2}{\rh_c c^2} + \frac{\rh_0 \Pi_0}{\rh_c c^2} = \Om_M + \Om_M \frac{\Pi_0}{c^2} = 1 \: . \ee In view of Eq. (23), the extra (dark) energy, needed to flatten the Universe, can be provided by the energy of the internal motions of a thermodynamically-involved-DM fluid.

On the other hand, upon consideration of Eq. (21), Eq. (19) is written in the form \be \left ( \frac{H}{H_0} \right )^2 = \left ( \frac{S_0}{S} \right )^3 \left [ 1 + 3 w \Om_M \ln \left ( \frac{S_0}{S} \right ) \right ] \: . \ee Equation (24) can be solved, explicitly, in terms of the error function (see Appendix A). However, it can become particularly transparent (and useful) if we take into account that, since $0 \leq w \leq 1$ and $\Om_M \simeq 0.3$, the combination $w \Om_M$ can be quite small, i.e., $w \Om_M \ll 1$. In this case, we may take the natural logarithm on both sides of Eq. (24), to obtain \be \ln \left ( \frac{H}{H_0} \right )^2 = \ln \left ( \frac{S_0}{S} \right )^3 + \ln \left [ 1 + w \Om_M \ln \left ( \frac{S_0}{S} \right )^3 \right ] . \ee Within the post-recombination era, $\frac{S_0}{S} \leq 1090$, hence $\ln \left ( \frac{S_0}{S} \right )^3 \leq 21$. Therefore, as long as $w \Om_M \ll 1$, we have \be \ln \left [ 1 + w \Om_M \ln \left ( \frac{S_0}{S} \right )^3 \right ] \simeq w \Om_M \ln \left ( \frac{S_0}{S} \right )^3 \: , \ee so that, for terms linear in $w \Om_M$, Eq. (24) results in \be H \simeq H_0 \left ( \frac{S_0}{S} \right )^{\frac{3}{2} (1 + w \Om_M)} \: . \ee In this case, using Eq. (12), we can solve Eq. (27), to determine the scale factor of the collisional-DM model (1), as follows \be S = S_0 \left ( \frac{\et}{\et_0} \right )^{\frac{2}{1 + 3 w \Om_M}} , \ee where we have defined the present-time value, $\et_0$, of the conformal time, $\et$, as \be \et_0 = \frac{2}{(1 + 3 w \Om_M) H_0 S_0} \: . \ee For $w \neq 0$, Eq. (28) is the natural generalization of the corresponding collisionless-DM model, the well-known Einstein-de Sitter (EdS) Universe $\left ( S \sim \et^2 \right )$ (see, e.g., Peacock 1999, pp. 77, 83 and 142 - 144). 

Eventually, in the collisional-DM model given by Eq. (1), the cosmological redshift parameter is defined as \be z + 1 = \frac{S_0}{S} \: , \ee thus Eq. (27) is written in the form \be H = H_0 ( 1 + z )^{\frac{3}{2} (1 + w \Om_M)} \: . \ee We note the striking functional similarity between Eq. (31) and the corresponding result for a dark-energy fluid with equation of state in the form of Eq. (3) (cf. Eqs. (13) and (14) of Perivolaropoulos 2007). In our case, however, $w \geq 0$ and, therefore, on the approach to $z = 0$, $H(z)$ decreases monotonically. In other words, a cosmological model filled with collisional DM, necessarily, decelerates its expansion.

This can be readily verified, by expressing the corresponding deceleration parameter, $q$, in terms of $H$ and $z$, as \be q (z) = \frac{dH / dz}{H(z)} (1+z) - 1 \ee (cf. Eq. (16) of Perivolaropoulos 2007), which, in view of Eq. (31), yields \be q (z) = \frac{1}{2} (1 + 3 w \Om_M) > 0 \: , \ee independently of $z$, even for $w = 0$. In other words, the model of a gravitating perfect-fluid source, as it stands, i.e., either pressureless (geodesic motions) or not (hydrodynamic flows), seems to be inappropriate for explaining the apparent accelerated expansion of the Universe. 

The actual reason is that it does not have to account for any acceleration at all. As we demonstrate in the next Section, in a Universe filled with collisional (i.e., thermodynamically-involved) DM, the observed dimming of the distant light-emitting sources can be explained without the assumption of the accelerated expansion.

\section{Mistreating the dark matter as collisionless}

When the (unexpected) dimming of the SNe Ia standard candles was first discovered, the common perception about the cosmos, to the best of our knowledge, was that the DM constituents are collisionless, thus the various motions in the Universe were (necessarily) interpreted as geodesic motions of test particles receding from each other, i.e., \be \frac{d \tilde{u}^{\kp} }{d \tilde{s}} + \tilde{\Gm}_{\mu \nu}^{\kp} \tilde{u}^{\mu} \tilde{u}^{\nu} = 0 \: . \ee Tilde variables are used, to distinguish the various quantities in Eq. (34) from the corresponding quantities used in Eq. (9), thus reflecting that the physical content of a collisionless-DM Universe (in which both the pressure and the energy of the internal motions are assumed to be negligible and, therefore, disregarded) is entirely different from that of the thermodynamically-involved-DM model (where $p, \: \Pi \neq 0$). 

In other words, the dynamical properties of a dust model are no longer described by $g_{\mu \nu}$, i.e., Eq. (1), but rather in terms of another metric tensor, $\tilde{g}_{\mu \nu}$, for which the corresponding (spatially-flat) line-element is written in the form \be d \tilde{s}^2 = R^2 (\et) \left [ c^2 d \et^2 - \left ( dx^2 + dy^2 + dz^2 \right ) \right ] \: . \ee Clearly, the evolution of this model is given in terms of the scale factor $R(\et)$. From the point of view of an observer who (mis)treats the DM as collisionless, $\tilde{g}_{\mu \nu}$ is the metric tensor upon which he/she should rely on, in interpreting observations.

However, we recall that, for this observer, the accumulated evidence in favour of spatial flatness (necessarily) leads to the assumption of an extra (dark) energy component, in contrast to the collisional-DM case, where this assumption would no longer be necessary. In the latter case, the appropriate candidate to provide the extra energy needed to flatten the Universe is already included in the model (the energy of the internal motions).

Furthermore, in the collisionless-DM scenario, every theoretical effort to interpret the (apparent) dimming of the SNe Ia standard candles, naturally, should also be based on $\tilde{g}_{\mu \nu}$ and the cosmologically relevant parameters arising from it. Accordingly, a possible explanation could be that, recently, the Universe accelerated its expansion (Riess et al. 1998; Perlmutter et al. 1999). This assumption, however, attributes unnecessarily-exotic properties to the extra amount of energy needed to account for the spatial flatness (e.g., it should be repulsive in nature, i.e., of negative pressure, etc.). Therefore, we cannot help but wondering whether there is another (more conventional) explanation to be found (also) within the context of the collisional-DM model.

In what follows, we demonstrate that both the observed dimming of the distant light-emitting sources and the accelerated expansion of the Universe could be only apparent, based on the misinterpretation of several cosmologically-relevant parameters, by someone who (although living in a Universe filled with collisional DM) insists on adopting the traditional (collisionless-DM) approach.

To explore this possibility, we note that the collisional-DM treatment of the Universe's matter content (in terms of which $p \neq 0$ and the motions of its constituents are, in principle, hydrodynamic flows) can be related to the collisionless-DM approach where $\tilde{p} = 0$ and the corresponding motions are, necessarily, geodesics, by means of a conformal transformation of the metric tensor (Kleidis \& Spyrou 2000; Spyrou 2005; Spyrou \& Tsagas 2004, 2010). Accordingly, from the original metric, $g_{\mu \nu}$, in terms of which the hydrodynamic flows have their well-known form given by Eq. (8), we can transfer the problem to a virtual metric, $\tilde{g}_{\mu \nu}$, in terms of which the volume elements of the fluid move along geodesic-like trajectories, i.e., their velocity-vector obeys Eqs. (34) (in connection, see Synge 1937; Lichnerowicz 1967, pp. 24 - 29 and 54 - 61; Carter 1979). The appropriate (conformal) transformation for such a transition is \be \tilde{g}_{\mu \nu} = F^2 (x^{\kp}) \: g_{\mu \nu}, \ee where, upon consideration of isentropic flows, the conformal factor $F (x^{\kp})$ takes on the functional form (Kleidis \& Spyrou 2000) \be F (x^{\kp}) = {\cal C} \left ( \frac{\varep + p}{\rh c^2} \right ) = {\cal C} \left [ 1 + \frac{1}{c^2} \left ( \Pi + \frac{p}{\rh} \right ) \right ] \: , \ee with ${\cal C}$ being an arbitrary (integration) constant. From Eq. (37), it becomes evident that $F (x^{\kp})$ is, essentially, the specific enthalpy of the ideal fluid under consideration. 

Verozub (2008) extrapolated these results to include every Riemannian space-time and not just the metric attributed to a bounded, perfect-fluid source. In particular, he showed that the adiabatic hydrodynamic motion of an ideal-fluid element in a space-time with metric tensor $g_{\mu \nu}$, takes place along the geodesic lines of a Riemannian manifold with metric tensor given by the combination of Eqs. (36) and (37).

With the aid of the technique developed by Kleidis \& Spyrou (2000), we then determine the scale factor of the spatially-flat cosmological model in Eq. (35), i.e., the scale factor of the Universe as inferred by someone who, although living in a collisional-DM Universe (where $p , \: \Pi > 0$ and $\frac{d p}{d \et} \neq 0$), misinterprets the DM as collisionless $(\tilde{p} = 0)$. 

In principle, one can (always) use a (conformal) transformation to remove (from the rhs of Eqs. (8)) either the pressure gradient, which measures the response to non-gravitational forces, or the pressure itself (see, e.g., Lichnerowicz 1967, p. 26). Nevertheless, something like this (usually) comes with a price (see, e.g., Bruneton \& Esposito-Farese 2007): The new metric, in terms of which the DM appears to be pressureless, is no longer a solution of general relativity (GR), but, rather, a solution to a modified theory of gravity. The reason is that, in terms of this new metric (i.e., after the transformation (36) is applied), the action of the original gravitational field is also modified, acquiring extra terms in addition to the Einstein-Hilbert Lagrangian (see Appendix B). In other words, every effort to treat a collisional-DM fluid as pressureless, cannot be accomplished in the context of GR. As a consequence, $R(\et)$ is no longer a solution of the original Friedmann equation, given by Eq. (11). 

In view of Eq. (36), the scale factor of the Universe as it is inferred by a supporter of the collisionless-DM scenario, $R(\et)$, is related to the corresponding quantity of the collisional-DM model, $S(\et)$, as follows \be R (\et) = F (x^{\kp}) S (\et) \: , \ee where, by virtue of Eqs. (7), (17), and (30), $F (x^{\kp})$ is given, in terms of $z$, by \be F(z) = \frac{\cal C}{\Om_M} \left ( 1 + w \Om_M \left [ 1 + 3 \ln (1 + z) \right ] \right ) \: . \ee In Eq. (39), the arbitrary integration constant, ${\cal C}$, can be determined, by demanding that, in the (isobaric) pressureless case, these two models should coincide, i.e., $\tilde{g}_{\mu \nu} = g_{\mu \nu}$. In other words, for $w = 0 = p$, $R(\et) = S(\et, w=0)$, which represents the EdS model. Hence, $F(w = 0) = 1$.

For $p = constant = p_0$, the first law of thermodynamics given in Eq. (13) yields \be \Pi + \frac{p_0}{\rh} = constant \: , \ee which, in the particular case of dust, where $p_0 = 0$, results in $\Pi = constant = \Pi_0$. Now, from Eq. (19) it becomes evident that, at the present epoch, $\Pi$ $(= \: \Pi_0)$ is (also) given by Eq. (21), even for $w = 0$. Accordingly, inserting Eq. (21) into Eq. (37), we find that, the condition $F(w = 0) = 1$ leads to \be {\cal C} = \Om_M \: . \ee As a consequence, Eq. (39) results in \be F(z) = 1 + w \Om_M \left [ 1 + 3 \ln (1 + z) \right ] \: . \ee Using Eqs. (38) and (42), we can then express several cosmologically relevant parameters of the collisional-DM model in terms of their collisionless-DM counterparts, such as the cosmological redshift, the luminosity distance, and the distance modulus of the various light-emitting sources, together with the Hubble and the deceleration parameters, which characterize the cosmic expansion.

\subsection{The cosmological redshift}

A supporter of the collisionless-DM scenario would define the corresponding cosmological redshift parameter, $\tilde{z}$, as \be \tilde{z} + 1 = \frac{R (\et_0)}{R (\et)} \: , \ee which, upon consideration of Eq. (38), is written in the form \be \tilde{z} + 1 = \frac{F (\et_0)}{F (\et)} (z + 1) \: , \ee where \be F (\et_0) = \frac{R(\et_0)}{S(\et_0)} = 1 + w \Om_M \: . \ee Taking into account Eq. (42), Eq. (44) results in \be \tilde{z} + 1 = \frac{1 + w \Om_M}{1 + w \Om_M [1 + 3 \ln (1 + z)]} (z + 1) \: . \ee Confining ourselves to relatively low values of the cosmological redshift parameter (e.g., $z < 5$), to ensure that the combination $3 w \Om_M \ln (1+z)$ remains sufficiently-lower than unity even for relatively large values of the combination $w \Om_M$ (e.g., $w \Om_M \sim 0.1$, i.e., $w \sim \frac{1}{3}$), we can apply the technique used in Eqs. (25) - (27), to obtain \be 1 + \tilde{z} \simeq (1 + z)^{1 - 3w \Om_M} \: . \ee In this case, we note that, for every (fixed) value of the cosmological redshift $z$, i.e., as defined in the collisional-DM model, the corresponding collisionless-DM quantity $\tilde{z}$ is always a little bit smaller $(\tilde{z} < z)$.

In other words, on observing a light-emitting source of the collisional-DM model, an observer who adopts the collisionless-DM scenario (realizing redshifts as $\tilde{z}$ instead of $z$), necessarily admits that this source lies a little bit farther $(z)$ than expected $(\tilde{z})$.

\subsection{The luminosity distance and the distance modulus}

Nowadays, the most direct and reliable method for determining, observationally, the (relatively) recent history of the Universe expansion, is to measure the redshift and the apparent luminosity (equivalently, the apparent magnitude, $m$) of cosmologically-distant indicators (standard candles), whose absolute luminosity (equivalently, the absolute magnitude, $M$) is assumed to be known.

SN Ia events constitute one of the most suitable cosmological standard candles (Plionis et al. 2009). With the aid of these events, a number of scientific groups have attempted to find evidence in support of a recently-accelerating stage of the Universe (Garnavich et al. 1998; Riess et al. 1998, 2001, 2004, 2007; Perlmutter et al. 1999; Knop et al. 2003; Tonry et al. 2003; Astier et al. 2006; Wood-Vasey et al. 2007; Kowalski et al. 2008; Hicken et al. 2009). In each and every one of these surveys, the SN Ia events, at peak luminosity, appear to be dimmer (i.e., they seem to lie farther away) than expected. This result was, eventually, accommodated within the context of the concordance model, by a DE fluid of negative pressure, with $\Om_{X} \sim 0.7$ (see, e.g., Spergel et al. 2003; Tonry et al. 2003). However, in view of Eq. (47), there may be also another, more conventional interpretation.

Photons travel along null geodesics, $d \tilde{s}^2 = 0 = ds^2$, which remain unaffected by conformal transformations. Accordingly, in both the collisional-DM and collisionless-DM approaches, the radial distance of a light-emitting source (in comoving coordinates) is the same, i.e., \be \tilde{r} = c \: (\et_r - \et_e) = r \: , \ee where $\et_r$ and $\et_e$ are the conformal times of reception and emission of light, respectively (usually, $\et_r = \et_0$).

In this case, with the aid of Eq. (47), the formula determining the luminosity distance in a spatially-flat collisional-DM model \be d_L (z) = r S(\et_0) (1 + z) \ee (see, e.g., Peacock 1999, p. 92) can be expressed in terms of the corresponding collisionless-DM quantity \be \tilde{d}_L (\tilde{z}) = \tilde{r} R(\et_0) (1 + \tilde{z}) \ee as \be \frac{d_L }{\tilde{d}_L} = \frac{1}{1 + w\Om_M} ( 1 + z )^{3 w \Om_M} \: . \ee This relation is very interesting. It suggests that, in a Universe containing collisional DM (i.e., as long as $w \neq 0$), there exists a characteristic (transition) value of the cosmological redshift, \be z_c = \left ( 1 + w \Om_M \right )^{\frac{1}{3w \Om_M}} - 1 \: , \ee such that, the luminosity distance of the various light-emitting sources located at $z > z_c$, is always larger than what is inferred by a supporter of the collisionless-DM scenario. Therefore, an observer who treats the DM as collisionless (measuring distances in terms of $\tilde{d}_L$) necessarily admits that any standard candle located at $z > z_c$ lies farther than expected, i.e., $\tilde{d}_L < d_L$. 

The same also happens in the case of the distance moduli corresponding to $d_L$ and $\tilde{d}_L$. The K-corrected distance modulus, $\mu (z) = m - M$, of a light-emitting source in the collisional-DM model is given by \be \mu (z) = 5 \log \left ( \frac{d_L}{Mpc} \right ) + 25 \ee (see, e.g., Narlikar 1983, Eqs. (13.10) and (13.12), p. 359), where $d_L$ is measured in megaparsecs $(Mpc)$. In a similar fashion, \be \tilde{\mu} (\tilde{z}) = 5 \log \left ( \frac{\tilde{ d}_L}{Mpc} \right ) + 25 \ee is the distance modulus of the same source, as defined by someone who, although living in the collisional-DM model, insists on adopting the (traditional) collisionless-DM approach. Subtracting Eqs. (53) and (54) by parts, and using Eq. (51), we obtain \be \mu = \tilde{\mu} + 15 w \Om_M \log ( 1 + z ) - 5 \log \left ( 1 + w \Om_M \right ) . \ee According to Eq. (55), any light-emitting source of the collisional-DM Universe located at $z > z_c$, from the point of view of an observer who insists on adopting the collisionless-DM approach (treating the various distance moduli in terms of $\tilde{\mu}$), appears to be dimmer than expected, i.e., $\tilde{\mu} < \mu$.

We cannot help but notice the prominent similarity between the characteristic value $z_c$ and the transition redshift, $z_t$, which, according to a supporter of the collisionless-DM scenario, signals the onset of the dimming of the SNe Ia standard candles, something that is interpreted (by such an observer) as an entry into a phase of accelerated expansion. As we see, what actually happens in a collisional-DM model is that, from the point of view of someone who incorrectly assumes the DM as collisionless (i.e., measuring cosmological distances in terms of $\tilde{d}_L$ instead of the truly-measured quantity $d_L$), an inflection point in the $\tilde{d}_L$ versus $z$ diagram (namely, $z_c$) will arise anyway. For $w\Om_M = 0.1$, i.e., $w = \frac{1}{3}$ (the DM consists of relativistic particles), the characteristic transition value in Eq. (52) is set at $z_c = 0.37$, while, for lower values of $w$, $z_c$ reaches up to $0.39$. These results lie within the observationally traced range of values concerning $z_t$, namely $z_t = 0.46 \pm 0.13$ (Riess et al. 2004, 2007). In other words, the spatially-flat collisional-DM model given by Eq. (28) does not suffer from the coincidence problem.

Therefore, if the DM possesses some sort of thermodynamical content, then, it is possible that \texttt{(i)} the "infernous" discrepancy between the expected value of the distance modulus $(\tilde{\mu})$ of a SN Ia standard candle and the corresponding observed one $(\mu)$, and \texttt{(ii)} the accompanying inflection point, $z_t$, that signals the transition from deceleration to acceleration, may both arise only because many cosmologists (although living in a collisional-DM model) insist instead on adopting the (traditional) collisionless-DM approach. In the next Section, we demonstrate that this is exactly what happens in a collisional-DM Universe. 

\subsection{The Hubble and the deceleration parameters}

By virtue of Eq. (38), the Hubble parameter inferred by a supporter of the collisionless-DM scenario, $\tilde{H}$, is written in terms of $H$ as \be \tilde{H} = \frac{1}{F} H - \frac{1}{S} \: \frac{d}{d \et} \left ( \frac{1}{F} \right ) \: , \ee from which, a much more interesting relation can be obtained in terms of the cosmological redshift. By taking into account that \be \frac{1}{S} \: \frac{d}{d \et} \left ( \frac{1}{F} \right ) = - (1 + z) \: H \: \frac{d}{dz} \left ( \frac{1}{F} \right ), \ee Eq. (56) is written in the form \be \tilde{H} = H \: \frac{d}{dz} \left ( \frac{1+z}{F} \right ), \ee which, in view of Eq. (42), results in \be \tilde{H} = H \: \frac{1 - 2 w \Om_M + 3 w \Om_M \ln (1 + z)}{(1 + w \Om_M [1 + 3 \ln (1 + z)])^2} \ee or else \bea \tilde{H} & = & H_0 (1 + z)^{\frac{3}{2} (1 + w \Om_M)} \nn \\ & \times & \frac{1 - 2 w \Om_M + 3 w \Om_M \ln (1 + z)}{(1 + w \Om_M [1 + 3 \ln (1 + z)])^2} \: , \eea where we have also used Eq. (31). We note that, to terms linear in $w \Om_M$, \be \tilde{H}_0 = H_0 (1 - 4 w \Om_M) \: , \ee i.e., within the context of the collisionless-DM approach, at the present epoch (when $z = 0$), the Universe expands only as long as $w \Om_M < \frac{1}{4}$, and, in any case, at a lower rate than the collisional-DM treatment $(H_0)$ implies. In view of Eq. (47), i.e., at relatively low values of $z$, Eq. (60) can be written in terms of $\tilde{z}$, as \bea \tilde{H} & = & H_0 \: (1 + \tilde{z})^{\frac{3 (1 + w \Om_M)}{2 (1 - 3 w \Om_M)}} \: (1 - 3 w \Om_M) \nn \\ & \times &  \frac{1 - 5 w \Om_M + 3 w \Om_M \ln (1 + \tilde{z}) + \textsl{O}(w \Om_M)^2}{ \left [ 1 - 2 w \Om_M + 3 w \Om_M \ln (1 + \tilde{z}) + \textsl{O}(w \Om_M)^2 \right ]^2} \: . \eea We verify that, to terms linear in $w \Om_M$, Eq. (61) is (also) valid at $\tilde{z} = 0$. 

By analogy with Eq. (32), a supporter of the collisionless-DM scenario would define the corresponding deceleration parameter, $\tilde{q}$, as \be \tilde{q} (\tilde{z}) = \frac{d \tilde{H}/d \tilde{z}}{\tilde{H} (\tilde{z})} ( 1 + \tilde{z}) - 1, \ee which, by virtue of Eq. (62), yields \be \tilde{q} (\tilde{z}) = \frac{1}{2} \left [ \frac{1 - 4 w \Om_M + 6 w \Om_M \ln (1 + \tilde{z}) + \textsl{O}(w \Om_M)^2}{1 - 10 w \Om_M + 6 w \Om_M \ln (1 + \tilde{z}) + \textsl{O}(w \Om_M)^2} \right ] . \ee Now, the condition for accelerated expansion $(\tilde{q} < 0)$ is translated to \bea &&[1 - 4 w \Om_M + 6 w \Om_M \ln (1 + \tilde{z})] \times \nn \\ &&[1 - 10 w \Om_M + 6 w \Om_M \ln (1 + \tilde{z})] < 0 \: , \eea from which, to terms linear in $w \Om_M$, we obtain \be \tilde{q} (\tilde{z}) < 0 \Leftrightarrow 1 - 14 w \Om_M + 12 w \Om_M \ln (1 + \tilde{z}) < 0 \: . \ee From Eq. (66) it becomes evident that, from the point of view of someone who insists on adopting the collisionless-DM approach, $\tilde{q} (\tilde{z}) < 0$ at cosmological redshifts \be \tilde{z} < \tilde{z}_t = e^{\frac{14 w \Om_M - 1}{12 w \Om_M}} - 1. \ee This relation is very interesting: It suggests that, if the Universe matter-content is treated as a collisional-DM fluid with $w$ being larger than a critical value, $w_c$, such that \be w \Om_M > w_c \Om_M = \frac{1}{14} \approx 0.0714 \ee (i.e., $w > w_c \approx 0.238$), then, from the point of view of someone who incorrectly assumes the DM as collisionless, there exists a transition value, $\tilde{z}_t$, of $\tilde{z}$, below which, the post-recombination Universe (as being realized by such an observer) is accelerating, independently of any notion of DE or the cosmological constant. 

In other words, if the Universe evolution is driven by a collisional-DM fluid with $w > w_c$, then, the apparent acceleration of the cosmic expansion could (very well) be due to a misinterpretation of several cosmologically-relevant parameters, by an observer who (although living in a cosmological model filled with collisional DM) insists on adopting the collisionless-DM approach. At the same time, for this observer, the cosmologically distant indicators would appear to be dimmer than expected (cf. Eq. (55)).

We recall here that the recent observational data concerning the SNe Ia standard candles set the transition redshift between accelerated and decelerated expansion at $z_t = 0.46 \pm 0.13$ (Riess et al. 2004). In this case, the combination of Eqs. (47) and (67) results in the non-linear algebraic equation involving the transition value, $z_t$, of the truly-measured quantity $z$ \be \left ( 1 + z_t \right ) \: e^{0.25/3 w \Om_M} = 3.2114 \left ( 1 + z_t \right )^{3 w \Om _M} \: . \ee Equation (69) can be solved numerically with respect to the combination $w \Om_M$. Accordingly, we verify that the value \be \left ( w \Om_M \right )_t = 0.0932 \pm 0.0060 \ee reproduces (exactly) the above observational result for $z_t$. 

By virtue of Eq. (70), we note that $w \simeq \frac {1}{3}$, i.e., compatibility of the collisional-DM approach with the observational data, currently available, suggests that, in Eq. (3), the pressure of the cosmic fluid under consideration is due to radiation. Nevertheless, in view of Eq. (17), the evolution of the rest-mass density indicates that, for every value of $w$, the spatially-flat model given by Eq. (28) is matter-dominated. Taken together, these results imply that the DM itself, being responsible for the non-vanishing pressure, consists of relativistic particles ("hot" DM). We verify this result by overplotting Eq. (55) in the Hubble ($\mu$ versus $z$) diagram of a SN Ia dataset. 

\subsection{Application to a sample of SN data}

An extended sample of 192 SN Ia events has been used by Davis et al. (2007) to scrutinize the viability of various DE scenarios. This sample\footnote{Available at \texttt{http://www.ctio.noao.edu/essence} or at \texttt{http://braeburn.pha.jhu.edu/$\sim$ariess/R06}} consists of 45 SNe from a nearby SN Ia dataset (Hamuy et al. 1996; Riess et al. 1999; Jha et al. 2006), 57 events from SNLS - the SuperNova Legacy Survey (Astier et al. 2006), 60 intermediate-redshift events from ESSENCE - the Equation of State: SupErNovae trace Cosmic Expansion program (Wood-Vasey et al. 2007), and 30 high-$z$ SNe from the Gold-07 sample (Riess et al. 2007).

To overplot Eq. (55) on the $\mu$ versus $z$ diagram of this dataset, first of all, we need to determine the functional form of the luminosity distance $\tilde{d}_L (\tilde{z})$ (and, through it, the corresponding form of the distance modulus, $\tilde{\mu} (\tilde{z})$), used by someone who, although living in a collisional-DM model, insists on the collisionless-DM approach. This observer performs calculations in the (traditional) framework of a pressureless Universe, adopting the corresponding formula of the luminosity distance. In a spatially-flat model, this formula is given by (see, e.g., Carroll et al. 1992) \be \tilde{d}_L (\tilde{z}) = \frac{2c}{\tilde{H}_0} \: \left ( 1 + \tilde{z} \right )^{1/2} \: \left [ \left ( 1 + \tilde{z} \right )^{1/2} - 1 \right ] \: , \ee representing the luminosity distance in the EdS Universe, the (conformally) pressureless counterpart of the collisional-DM model given by Eq. (28). 

However, we need to stress that, in a collisional-DM Universe, the measured quantity (corresponding to the cosmological redshift) is $z$ and not $\tilde{z}$, as it is (falsely) admitted by someone who (mis)treats the DM as collisionless. Therefore, in order to include, also, the function $\tilde{\mu} (\tilde{z})$ in the Hubble diagram of the SN Ia dataset used by Davis et al. (2007), we have to express $\tilde{d}_L (\tilde{z})$ in terms of the truly-measured quantity, $z$. It can be done (appropriately) by inserting Eqs. (47) and (61) into Eq. (71), to obtain \bea \tilde{d}_L (z) & = & \frac{2c}{\left ( 1 - 4w \Om_M \right ) H_0} \: (1 + z)^{\frac{1}{2}(1 - 3w\Om_M)} \nn \\ & \times & \left [ ( 1 + z)^{\frac{1}{2}(1 - 3w\Om_M)} - 1 \right ] \: . \eea However, as we have already mentioned, this is not the case for a supporter of the collisionless-DM scenario. In depicting Eq. (54) - with $\tilde{d}_L (\tilde{z})$ being given by Eq. (71) - on the $\mu$ versus $z$ diagram of a sample of SN events, this observer (admitting that $w = 0$), unavoidably, misinterprets the measured quantity $z$ as $\tilde{z}$ (and the quantity $H_0$ as $\tilde{H}_0$). In other words, the theoretical formula of the luminosity distance that is used by someone who, although living in a (spatially-flat) collisional-DM model, insists on adopting the collisionless-DM approach is (incorrectly) written in the form of Eq. (71) with $\tilde{z}$ simply replaced by $z$ \be \tilde{d}_L (z) = \frac{2c}{H_0} \: \left ( 1 + z \right )^{1/2} \: \left [ \left ( 1 + z \right )^{1/2} - 1 \right ] \: , \ee instead of that given by Eq. (72). In what follows, we admit that $H_0 = 70.5 \: Km/sec/Mpc$ (Komatsu et al. 2009), hence $2 c/H_0 = 8509.8 \: Mpc$.

We then overplotted in the Hubble diagram of the SN Ia dataset used by Davis et al. (2007) the theoretical curves corresponding to the distance moduli: $\mu (z)$ for $w \Om_M = 0.16$ (green solid line); $\tilde{\mu} (z)$ (also for $w \Om_M = 0.16$) with $\tilde{d}_L (z)$ being given by Eq. (72) (orange solid line); and $\tilde{\mu} (z)$ with $\tilde{d}_L (z)$ being given by Eq. (73) (dashed line). The outcome is presented in Fig. 1. We observe that, the (appropriately translated in terms of $z$) collisionless-DM quantity $\tilde{\mu} (z)$ - with $\tilde{d}_L (z)$ being given by Eq. (72) (orange solid line) is quite far from being able to reproduce these data, although, for $z \leq 1.75$, it is much closer to the "real world" (the $\mu$ versus $z$ distribution of the SN Ia data available) than the incorrectly used quantity $\tilde{\mu} (z)$ - with $\tilde{d}_L (z)$ being given by Eq. (73) (dashed line). 

The situation changes, completely, when someone takes into account the thermodynamical content of a collisional-DM fluid with $w \Om_M = 0.16$, thus using Eq. (55) instead of Eq. (54) alone. In this case, the function $\mu (z)$ (green solid line) seems to fit the entire dataset under consideration quite accurately. 

\begin{figure}[ht!]
\centerline{\mbox {\epsfxsize=9.cm \epsfysize=7.cm
\epsfbox{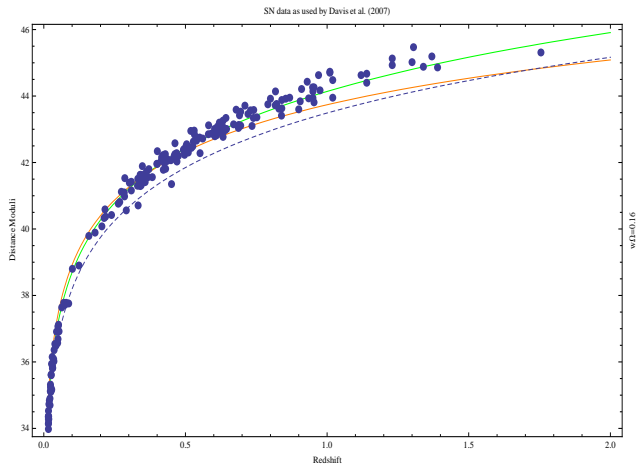}}} \caption{Hubble diagram of the SN Ia sample used by Davis et al. (2007). Overplotted are the theoretical curves, corresponding to the distance moduli: $\mu (z)$ for $w \Om_M = 0.16$ (green solid line); $\tilde{\mu} (z)$ (also for $w \Om_M = 0.16$) with $\tilde{d}_L (z)$ being given by Eq. (72) (orange solid line); and $\tilde{\mu} (z)$ with $\tilde{d}_L (z)$ being given by Eq. (73) (dashed line). We observe that, after the thermodynamical content of a collisional-DM fluid is taken into account, the theoretical curve representing the distance modulus, $\mu (z)$ (Eq. (55)), fits the entire dataset quite accurately (green line).}
\end{figure}

As we observe in Fig. 2, apart from a small number of SN events (4 of 192), the entire dataset used by Davis et al. (2007) lies within the stripe formed by the theoretical curves $\mu (z)$, corresponding to $w \Om_M = 0.10$ (red solid line) and $w \Om_M = 0.19$ (blue solid line), while the best fit to this sample appears to be achieved for $w \Om_M = 0.16$ (green line). In a spatially-flat cosmological model, it is clear that, the collisional-DM treatment is not only much closer to, but, actually, can reproduce quite accurately the $\mu$ versus $z$ distribution of the SN Ia sample used by Davis et al. (2007). 

Nevertheless, there is a "delicate point" in this treatment. For $w \Om_M \geq 0.10$, we have $w \geq \frac{1}{3}$. In other words, compatibility of the collisional-DM approach with the currently available observational data, suggests that the cosmic fluid (i.e., above all, the DM itself) consists of relativistic particles.

\begin{figure}[ht!]
\centerline{\mbox {\epsfxsize=9.cm \epsfysize=7.cm
\epsfbox{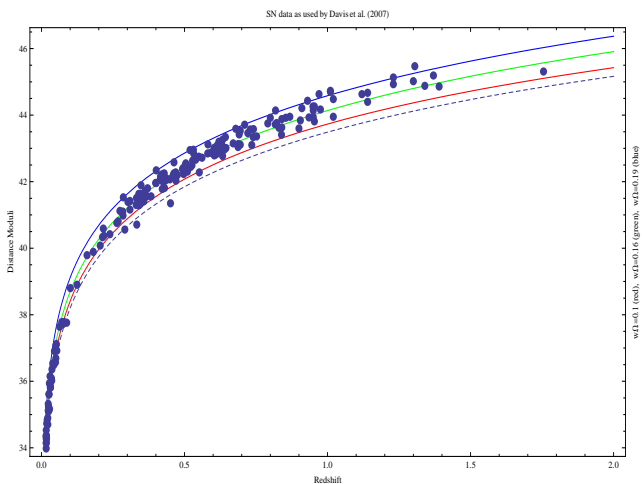}}} \caption{Overplotted in the Hubble diagram of the SN Ia sample used by Davis et al. (2007), are the theoretical curves of the distance modulus in the collisional-DM model, i.e., Eq. (55), for several values of the combination $w \Om_M$. Almost the entire dataset lies within the stripe formed by the curves $\mu (z)$, corresponding to $w \Om_M = 0.10$ (red solid line) and $w \Om_M = 0.19$ (blue solid line), while, the {\em best fit} to this sample is achieved for $w \Om_M = 0.16$ (green line). Once again, the dashed line represents the (incorrectly used) theoretical curve $\tilde{\mu} (z)$ with $\tilde{d}_L (z)$ being given by Eq. (73).}
\end{figure}

Taken together, these results strongly suggest that, if the DM constitutes a perfect fluid (of relativistic particles) with thermodynamical content, then, what is interpreted as the "dimming" of the SNe Ia standard candles might only be apparent, provided that the cosmologists no longer insist on adopting the collisionless-DM approach.

In other words, in a Universe filled with collisional (and relativistic) DM, the unexpected dimming of the distant light-emitting sources can be explained in a more conventional way, than that implemented within the context of the accelerated expansion. Hence, before inventing any new theory, it is useful to allow for a suitable use of the (so far) neglected degrees of freedom (energy of the internal motions, pressure, etc.). As we have shown, these internal physical characteristics can reveal their influence on several parameters of cosmological significance (scale factor, cosmological redshift, luminosity distance, Hubble and deceleration parameters) and yield a consistent alternative to the currently-accepted DE concept. 

\subsection{An independent confirmation from BAO's data analysis}

Among the various (currently available) techniques for tracing the expansion history of the Universe, baryon acoustic oscillations (BAO) appear to have the lowest level of systematic uncertainty (Albrecht et al. 2006).

BAO is a series of peaks and troughs, with wavenumber (approximately) $0.06 h$ $\left ( Mpc \right )^{-1}$ (Eisenstein et al. 2005), arising, on large scales, in the power spectrum of matter fluctuations after recombination. They occur because the primordial cosmological perturbations excite sound waves in the relativistic plasma of the early Universe (Silk 1968; Peebles \& Yu 1970; Sunyaev \& Zel'dovich 1970; Bond \& Efstathiou 1984, 1987; Holtzman 1989; Hu \& Sugiyama 1996). This process continues for a short time-interval after recombination, until the epoch known as baryon-drag epoch $(z_d \simeq 1089)$, when, eventually, the baryons are released from the Compton "drag" of photons (Eisenstein \& Hu 1998).

The BAO's wavenumber is related to the comoving distance of the sound horizon at the baryon-drag epoch, $r_s$, which depends on both the total-mass density of the Universe and the corresponding baryonic quantity (see, e.g., Wang 2006). WMAP constraints on $\Om_M h^2$ and $\Om_B h^2$ (Komatsu et al. 2009) suggest that $r_s (z_d) \simeq 153.5 \; Mpc$. In other words, BAO are met on relatively-large scales, which, at the present epoch, are (still) in the linear regime. It is, therefore, expected that, these acoustic signatures should be present (also) in the distribution of galaxies (Goldberg \& Strauss 1998; Meiksin, White \& Peacock 1999; Seo \& Eisenstein 2005; Eisenstein, Seo \& White 2007). 

Since the comoving distance of the sound horizon at the baryon-drag epoch is known, we can use BAO as standard rulers, to determine the functional form of the distance - redshift relation in the Universe. In particular, by determining observationally the apparent size of a BAO's peak (it is identified as a clustering within a galaxy distribution), we can extract reliable estimates of the Hubble parameter, as well as the angular diameter distance, at the cosmological redshift, in which, this acoustic signature is observed (Blake \& Glazebrook 2003; Hu \& Haiman 2003; Seo \& Eisenstein 2003).

The spectroscopic Sloan Digital Sky Survey (SDSS) Data Release 7 (DR7) represents the final set of galaxies, observed using the original SDSS-target selection criteria (York et al. 2000). When combined with the data released from the 2-degree Field (2dF) galaxy-redshift survey (Colless et al. 2003), the resulting sample comprises $893 \: 319$ galaxies and covers a solid angle of $9100$ $deg^2$. An analysis of the clustering of galaxies within this sample, inferred BAO signals in the power spectrum measured in several slices of the cosmological redshift (Percival et al. 2010).

To minimize the systematic errors arising from treating the line-of-sight dilation in an equivalent manner to the transverse one (a common bias, involving the so-called Alcock-Paczy\'{n}ski (1979) effect), cosmologists defined the distance measure \be D_V (Z) \equiv \sqrt[3] {r_{\vert \vert} r_{\bot}^2} = \left [ (1+Z)^2 D_A^2 (Z) \frac{cZ}{H(Z)} \right ]^{1/3} \ee (Eisenstein et al. 2005; Percival et al. 2007), where $Z$ is the observationally-determined (i.e., spectroscopically-measured) value of the cosmological redshift of a large-scale structure, and \be D_A (Z) = \frac{S (\et_0) \:  r}{1+Z} \ee is the corresponding angular diameter distance in a spatially-flat cosmological model (see, e.g., Peacock 1999, p. 22). The combination of Eqs. (74) and (75) results in \be D_V (Z) = \left [ S^2 (\et_0) \: r^2 \frac{cZ}{H(Z)} \right ]^{1/3} , \ee where, once again, $r$ is the radial distance (in comoving coordinates) of the structure formation under consideration.

According to Percival et al. (2010), there exists a robust, statistically-independent (distance) constraint, arising from BAO's data, which involves the value of the ratio \be f = \frac{D_V (0.35)}{D_V (0.2)} = 1.736 \pm 0.065 \: . \ee Clearly, in a collisional-DM model, Eq. (77) represents the truly-measured value of $f$. In this case, we ask ourselves what is observed by someone who, although living in a collisional-DM model, insists on adopting the (traditional) collisionless-DM approach. 

To answer this question, we note that, when, either the collisional-DM-oriented observer or the corresponding collisionless-DM one, spectroscopically measures the cosmological redshift of a particular large-scale structure, they both refer to the same quantity, $Z$. Accordingly, a supporter of the collisionless-DM scenario would express Eq. (76) in the form \be \tilde{D}_V (Z) = \left [ R^2 (\et_0) \: r^2 \frac{c Z}{\tilde{H}(Z)} \right ]^{1/3} , \ee where we have also used Eq. (48). Equation (78) differs from Eq. (76) only in the definition of the present-time value of the scale factor and in the functional dependence of the Hubble parameter on $Z$. In this case, upon consideration of Eqs. (45) and (60), we obtain \bea &&\frac{\tilde{D}_V (Z)}{D_V (Z)} = \left ( 1 + w \Om_M \right )^{2/3} \left ( \frac{H}{\tilde{H}} \right )^{1/3} \; \Rarrow \\ &&\frac{\tilde{D}_V (Z)}{D_V (Z)} = \left ( 1 + w \Om_M \right )^{2/3} \frac{\left [ 1 + w \Om_M + 3w \Om _M \ln (1+Z) \right ]^{2/3}}{\left [ 1 - 2w \Om_M + 3w \Om _M \ln (1+Z) \right ]^{1/3}} . \nn \eea Now, as far as a supporter of the collisionless-DM scenario is concerned, the BAO's constraint given by Eq. (77) is translated as \bea \tilde{f} = \frac{\tilde{D}_V (0.35)}{\tilde{D}_V (0.2)} & = & \left [ \frac{1 - 2w \Om_M +3w \Om_M \ln (1.2)}{1 - 2w \Om_M +3w \Om_M \ln (1.35)} \right ]^{1/3} \nn \\ & \times & \left [ \frac{1 + w \Om_M +3w \Om_M \ln (1.35)}{1 + w \Om_M +3w \Om_M \ln (1.2)} \right ]^{2/3} \nn \\ & \times & \frac{D_V (0.35)}{D_V (0.2)} \: , \eea which for $w \Om_M = 0.10$ results in \be \tilde{f} \vert_{w \Om_M = 0.1} = \frac{\tilde{D}_V (0.35)}{\tilde{D}_V (0.2)} = 1.766 \: , \ee and for $w \Om_M = 0.16$, we have that \be \tilde{f} \vert_{w \Om_M = 0.16} = \frac{\tilde{D}_V (0.35)}{\tilde{D}_V (0.2)} = 1.788 \: . \ee Both values of $\tilde{f}$ lie within the range of values $[1.671 \: , \: 1.801]$ of the measured quantity, $f$, given by Eq. (77). Therefore, as far as a supporter of the collisionless-DM scenario is concerned, the observational constraint, $\tilde{f}$, inferred from BAO's data, is (at least) compatible with (if not almost identical to) the corresponding constraint, $f$, obtained within the context of the collisional-DM model.

However, when the collisionless-DM-oriented observer attempts to verify Eq. (81) (or Eq. (82)) theoretically, a controversy arises. This observer applies Eq. (78) to the EdS Universe (the pressureless counterpart of the collisional-DM model under study), which accommodates the standard ($w = 0$) cold dark matter $(SCDM)$ cosmology. In this case, \be \tilde{D}_V (Z) = 2^{2/3} \frac{c}{\tilde{H}_0} Z^{1/3} \frac{\left [ (1+Z)^{1/2} - 1 \right ]^{2/3}}{(1 + Z)^{5/6}} , \ee hence, a supporter of the collisionless-DM scenario obtains \be \tilde{f}_{SCDM} = \frac{\tilde{D}_V (0.35)}{\tilde{D}_V (0.2)} = 1.553 \ee (see, also, Percival et al. 2010). Clearly, there is a difference between the theoretical result of Eq. (84) and the observational one given by Eq. (81) (or Eq. (82)). As far as the supporter of the collisionless-DM scenario is concerned, one way to compensate this difference, is to impose the existence of an extra (dark) energy component or the cosmological constant. However, even when he/she does so, i.e., within the context of the $\Lm CDM$ model, the theoretical value of the distance constraint, $\tilde{f}$, induced by the signature of BAO on cosmic structure, reaches up to \be \tilde{f}_{\Lm CDM} = 1.670 \ee (Eisenstein et al 2005; Percival et al. 2010). Although marginally, $\tilde{f}_{\Lm CDM}$ lies outside the range of values of the truly-measured quantity, $f$, given by Eq. (77). In view of the above-mentioned results, we may conclude that, within the context of the collisional-DM approach, the BAO's-oriented observational data require (and acquire) a more sophisticated interpretation than the one provided by the (collisionless) $\Lm CDM$ model. 

\section{Discussion}

We have examined the possibility that the extra (dark) energy needed to flatten the Universe is represented by the energy of the internal motions of a collisional-DM fluid. Accordingly, we have considered the evolution of a cosmological (toy-) model driven by a gravitating fluid (consisting of DM - dominant - and baryonic matter - subdominant) with thermodynamical content. As a consequence, the energy of this fluid's internal motions has also been taken into account as a source of the universal gravitational field. Accordingly, we have asked ourselves, whether this model can also accommodate the apparent dimming of the cosmologically distant indicators and the associated phase of accelerated expansion.

In particular, since observational data indicates that WIMPs (of which the DM is believed to consist) can be collisional, we have assumed that the matter of the Universe (although resembling test particles receding from each other) can be represented by the volume elements of a (classical) collisional-DM fluid with some sort of internal structure, hence thermodynamical content. In this way, we have been able to determine the "correct" form of the scale factor, which (under the assumption that the DM is thermodynamically involved) governs the evolution of the Universe (modeled as a spatially-flat RW space-time), in addition to a series of parameters of cosmological significance. 

Our findings are quite promising. In principle, the energy of the internal motions of the collisional-DM fluid can account for the (extra) DE, so that, at the present epoch, $\Om = 1$ (cf. Eq. (23)), while the post-recombination Universe remains ever-decelerated (cf. Eq. (33)).

We next attempted to determine what is inferred by someone who, although living in a collisional-DM model, insists on adopting the (traditional) collisionless-DM approach.

To do so, we have applied the technique developed by Kleidis \& Spyrou (2000). With the aid of this technique, we have derived the (conformal) transformation (cf. Eqs. (38) and (42)), which relates the collisional-DM description of a cosmological model (in terms of which $p, \: \Pi > 0$ and $\frac{d p}{d \et} \neq 0$) to the corresponding collisionless-DM (pressureless) approach. With such a "tool" at hand, we have explored the way that a supporter of the collisionless-DM scenario interprets observations carried out in a collisional-DM Universe. In passing, we note that, within the context of general relativity, every effort to treat a collisional-DM model as pressureless is questioned (Appendix B). 

The debate between collisional- and collisionless-DM approach is, definitely, in favor of the former. In particular, for every value of the cosmological redshift $(z)$, as it is defined in the collisional-DM model, the corresponding collisionless-DM quantity, $\tilde{z}$, is always a little bit smaller (cf. Eq. (47)). As a consequence, in the collisional-DM model there is a characteristic value of the cosmological redshift, $z_c$ (cf. Eq. (52)), above which, the luminosity distance of the various light-emitting sources is always higher than what is inferred by an observer who treats the DM as pressureless (cf. Eq. (51)). In other words, from the point of view of someone who (although living in a collisional-DM model) insists on adopting the (traditional) collisionless-DM approach, the cosmologically-distant indicators, located at $z > z_c$, seem to lie farther away (i.e., appear dimmer) than expected (cf., also, Eq. (55)).

The similarity between the characteristic value $z_c$ and the (observationally-traced) transition redshift, $z_t$, which, according to a supporter of the collisionless-DM scenario, signals the onset of the dimming of the SNe Ia standard candles, is obvious. 

On the other hand, after the thermodynamical content of a collisional-DM fluid is taken into account, the theoretical curve representing the distance modulus, $\mu (z)$ (now given by Eq. (55)), fits the Hubble diagram of an extended sample of SN Ia standard candles quite accurately (green solid line in Fig. 1), in contrast to the corresponding collisionless-DM quantity, $\tilde{\mu} (\tilde{z})$, given either (appropriately) by the combination of Eqs. (54) and (72) (orange solid line in Fig. 1) or (incorrectly) by the combination of Eqs. (54) and (73) (dashed line in Fig. 1). In fact, a cosmological model filled with collisional DM could accommodate the majority of the currently-available observational data, including, also, those related to BAO, without the need for any dark energy or the cosmological constant (cf. Eqs. (81) and/or (82)).

At the same time, from the point of view of an observer who treats the DM as collisionless, the Universe appears to be either accelerating or decelerating, depending on the value of the cosmological redshift (cf. Eq. (67)).

In this case, the quantity $w$, which, in the collisional-DM approach, parameterizes the various flows, also plays another (more interesting) role. As we have found, for $w \Om_M \geq 0.0714$, there exists a (theoretically-determined) transition value, $\tilde{z}_t$, of the (collisionless-DM-oriented) cosmological redshift, $\tilde{z}$, such that, for $\tilde{z} < \tilde{z}_t$, we have $\tilde{q} < 0$, i.e., from the point of view of someone who adopts the (traditional) collisionless-DM approach, the Universe is accelerating, without the need for any DE or the cosmological constant. Accordingly, taking into account the observational result that the transition redshift between accelerated and decelerated expansion is set at the value $z_t = 0.46 \pm 0.13$ of the truly measured quantity $z$, we have determined the precise value of the combination $w \Om_M$, for which the collisional-DM approach to the post-recombination Universe is compatible with observations, i.e., $ \left ( w \Om_M \right )_t = 0.0932 \pm 0.0060$. This result is in complete agreement with what is implied by Fig. 2, i.e., $w \simeq \frac{1}{3}$. In other words, compatibility of the collisional-DM treatment under study with the observational data currently available, suggests that the DM itself consists of relativistic particles (hot DM). 

In conclusion, the assumption that the DM constituents can be both collisional and relativistic, could provide a reasonable and conventional explanation for several open aspects of modern cosmology, including: \texttt{(i)} The extra (dark) energy needed to flatten the Universe, which can be compensated by the energy of the internal motions of the collisional-DM fluid. \texttt{(ii)} The observed dimming of the SNe Ia standard candles and the apparent accelerated expansion of the Universe, both of which might be due to the misinterpretation of several cosmologically relevant parameters by those observers who, although living in a collisional-DM Universe, insist on adopting the collisionless-DM approach. The absence of any ad hoc fine-tuning in our study, makes the above (purely theoretical) results very promising.

In spite of all the above advantages of our model, we have to point out that this is, definitely, a toy model. The collisional-DM approach, developed in this article, can (and should) be debated, at least, along the following lines: 

\texttt{(i)} The collisional-DM treatment of the Universe does not alleviate the {\em age problem}, but, rather, makes it harder. In the model given by Eq. (28), the coordinate time, $t$, is related to the corresponding conformal quantity $(\et)$ by \be t = \int_0^{\et} S(\et) d \et = \frac{1}{1 + w \Om_M} \left ( \frac{2}{3 H_0} \right ) \left ( \frac{\et}{\et_0} \right )^{\frac{3 (1 + w \Om_M) }{1 + 3 w \Om_M}} \: . \ee Admitting that, at the present epoch (when $t = t_0$ and $\et = \et_0$) $H_0 \approx 70.5 \; Km/sec/Mpc$ (Komatsu et al. 2009), we find that, within the context of the model given by Eq. (28), the age of the Universe is \be t_0 = \frac{1}{1 + w \Om_M} \times 9.3 \; Gys \: , \ee i.e., less than $9.3$ billion years. Clearly, Eq. (87) could be a serious drawback of the spatially-flat collisional-DM model. 

\texttt{(ii)} The compatibility of the collisional-DM approach with the observational data currently available (cf. Eq. (70) and/or Fig. 2), suggests that the matter content of the dark sector consists of hot DM. For the time being, the conventional theory of hot DM does not appear to conform with the large-scale structure of the Universe (see, e.g., Hooper 2009), although there are recent studies that appear to challenge this result (see, e.g., Farrar \& Peebles 2004; Gubser \& Peebles 2004). Nevertheless, a hot-DM model looks much less {\em exotic} than most of the (currently-investigated) DE scenarios.

In any case, the assumption that the Universe matter content (basically its DM component) can be collisional (in the sense that it also possesses some sort of thermodynamical content), is to be seen as a natural effort to take into account all the (so far, practically, neglected) internal physical characteristics of a classical cosmological fluid as sources of the universal gravitational field. 

As we have shown, under this assumption, one can compensate for the majority of the recent observational data, inferring that $\Om \simeq 1$, as well as the "unexpected" dimming of the SNe Ia standard candles and the "apparent" accelerated expansion of the Universe, not to mention the compatibility with BAO's data analysis, without the need for any DE or the cosmological constant, and (certainly) without suffering from the coincidence problem. From the point of view of someone who (although living in the collisional-DM model) insists on adopting the collisionless-DM approach, an inflection point (in the $\tilde{d}_L$ versus $z$ diagram) arises, anyway, at relatively low values of the cosmological redshift.

Although speculative, the idea that the DE (needed to flatten the Universe) could be attributed to the internal motions of a collisional-DM fluid, is (at least) intriguing and should be explored further and scrutinized in the search for conventional alternatives to the DE concept. 

\begin{acknowledgements}

The authors would like to thank Professors George F. R. Ellis and Manolis Plionis, together with Drs. Leandros Perivolaropoulos and Christos G. Tsagas, for their constructive critisism, many illuminating discussions and useful comments on an earlier version of this article, as well as on this kind of research, in general. We also thank the anonymous referee for his/her kind comments and his/her useful suggestions, which improved this paper's final form. Finally, financial support by the Research Committe of the Technological Education Institute of Serres, under grant SAT/ME/151210-71/01, is gratefully acknowledged. 

\end{acknowledgements}

\section*{Appendix A}

In this Appendix, we present the exact solution to Eq. (24) in the text and its subsequent reduction to Eq. (28). In paricular, upon consideration of Eq. (12), Eq. (24) reads $$ \left ( \frac{S^{\prime}}{S^2} \right )^2 = H_0^2 \: \left ( \frac{S_0}{S} \right )^3 \left [ 1 + 3 w \Om_M \ln \left ( \frac{S_0}{S} \right ) \right ] \eqno{(A1)} $$ and can be cast in the (more convenient) form $$ \left [ \left ( \sqrt { \frac{S}{S_0}} \right )^{\prime} \right ]^2 = \left ( \frac{H_0 S_0}{2} \right )^2 \left [ 1 + 3 w \Om_M \ln \left ( \frac{S_0}{S} \right ) \right ]. \eqno{(A2)} $$ In an expanding Universe, Eq. (A2) is valid for every $\et$ within the past light-cone (where $S < S_0$) and it can be solved in terms of the error function (see, e.g., Abramowitz \& Stegun 1970, p. 297) as $$ \frac{\sqrt{\pi}}{\sqrt{6 w \Om_M}} e^{1/6 w \Om_M} \left [ erf \left ( \sqrt { \frac{1 + 3 w \Om_M \ln \left ( \frac{S_0}{S} \right )}{6 w \Om_M}} \right ) - 1 \right ] $$ $$ = \pm \left ( 1 + 3 w \Om_M \right ) \frac{H_0 S_0}{2} \: \et. \eqno{(A3)} $$ Equation (A3) becomes particularly transparent (and useful) if we restrict ourselves to the limiting case, where $w \Om_M \ll 1$.

For $w \Om_M \ll 1$, the argument of the error function in Eq. (A3) becomes very large. In this case, the special function under consideration behaves as $$ erf (\varphi) \simeq 1 - \frac{1}{\sqrt{\pi}} \: \frac{1}{\varphi} e^{- \varphi^2} \eqno{(A4)}$$ (see, e.g., Lebedev, Eq. (2.2.4), p. 19) and, therefore, Eq. (A3) results in $$ \frac{ \left ( \frac{S}{S_0} \right ) }{1 + 3 w \Om_M \ln \left ( \frac{S_0}{S} \right )} = \left ( 1 + 3 w \Om_M \right )^2 \left ( \frac{H_0 S_0}{2} \right )^2 \: \et^2. \eqno{(A5)}$$ We now take the natural logarithm on both sides of Eq. (A5), to obtain $$ \ln \left ( \frac{S}{S_0} \right ) - \ln \left [ 1 + w \Om_M \ln \left ( \frac{S_0}{S} \right )^3 \right ] \nn $$ $$= \ln \left [ \left ( 1 + 3 w \Om_M \right )^2 \left ( \frac{H_0 S_0}{2} \right )^2 \: \et^2 \right ]. \eqno{(A6)} $$ Within the post-recombination era, $\frac{S_0}{S} \leq 1090$, hence $\ln \left ( \frac{S_0}{S} \right )^3 \leq 21$. Therefore, as long as $w \Om_M \ll 1$, Eq. (26) holds, hence Eq. (A6) is written in the form $$ \left ( \frac{S}{S_0} \right )^{1 + 3 w \Om_M} = \left (1 + 3 w \Om_M \right )^2 \left ( \frac{H_0 S_0}{2} \right )^2 \: \et^2 \: , \eqno{(A7)}$$ which, in view of the definition given by Eq. (29), results in $$ S = S_0 \left ( \frac{\et}{\et_0} \right )^{\frac{2}{1 + 3 w \Om_M}} , \eqno{(A8)} $$ that is, Eq. (28).

\section*{Appendix B}

In curved space-time, the Einstein-Hilbert (EH) action, which governs the dynamical evolution of the gravitational field, $g_{\mu \nu}$, within the context of GR, is given by (see, e.g., Papapetrou 1973, Eq. (33.29), p. 122) $${\cal I}_{EH} = \int \sqrt{-g} \: \left ( {\cal R} + \kp {\cal T} \right ) d^4 x \: , \eqno{(B1)}$$ where $g$ is the determinant of the metric tensor $g_{\mu \nu}$, ${\cal R}$ is the scalar curvature, $\kp = 8 \pi G/c^4$, and ${\cal T}$ is the {\em trace} of the energy-momentum tensor of the Universe matter-energy content. In a similar fashion, the action of the gravitational field in terms of $\tilde{g}_{\mu \nu}$, i.e., after the transformation indicated in Eq. (36) is performed, is written in the form $$\tilde{{\cal I}} = \int \sqrt{-\tilde{g}} \: \left ( \tilde{{\cal R}} + \kp \tilde{{\cal T}} \right ) d^4 x \: , \eqno{(B2)}$$ which, upon consideration of Eqs. (36) and (39) of Kleidis \& Spyrou (2000), results in $$\tilde{{\cal I}} = {\cal I}_{EH} - 12 \int \sqrt{-g} \; \frac{\Dl F}{F} \: d^4 x \: , \eqno{(B3)}$$ where $\Dl F = g^{\mu \nu} F_{; \: \mu \nu}$ is the d' Alembert operator with respect to the original metric, $g_{\mu \nu}$. It is evident that, every effort to express $\tilde{{\cal I}}$ in terms of the original gravitational field, $g_{\mu \nu}$, yields the appearence of extra terms in the gravitational action, in addition to the EH one. In this way, variation of Eq. (B3) with respect to $g^{\mu \nu}$ does not lead to the Einstein field equations of GR (i.e., those derived from Eq. (B1)), thus (necessarily) resulting in a modified theory of gravity (in connection, see, e.g., Bruneton \& Esposito-Farese 2007).

\end{document}